\documentstyle[epsf]{europhys}

\def\stars{\bigskip\centerline{***}\medskip}

\newif\ifboo \boofalse

\newcommand{\mxi}{\mbox{\boldmath{$\xi$}}}
\newcommand{\mzeta}{\mbox{\boldmath{$\zeta$}}}

\newcommand{\A}{\mbox{{\boldmath $A$}}}
\newcommand{\G}{\mbox{{\boldmath $G$}}}

\newcommand{\J}{{\mbox{{\boldmath $J$}}}}

\newcommand{\mz}{{\mbox{{\boldmath $z$}}}}

\newcommand{\n}{{\mbox{{\boldmath $n$}}}}

\newcommand{\PP}{{\mbox{{\boldmath $P$}}}}

\begin{document}

\euro{}{}{}{}
\Date{}
\shorttitle{K. Nakamura {\em et al} STATISTICAL MECHANICS OF $GF(q)$ CODES}

\title{Statistical Mechanics of Low-Density Parity Check 
Error-Correcting Codes over Galois Fields}
\author{Kazutaka Nakamura$^{1}$\footnote{knakamur@fe.dis.titech.ac.jp},
Yoshiyuki Kabashima$^{1}$\footnote{kaba@dis.titech.ac.jp} and
David Saad$^{2}$\footnote{saadd@aston.ac.uk}}
\institute{$^{1}$Department of Computational Intelligence and Systems Science,
Tokyo Institute of Technology, Yokohama 2268502, Japan. \\
$^{2}$The Neural Computing Research Group, Aston
University, Birmingham B4 7ET, UK.\\}
\rec{}{}
\pacs{\Pacs{89}{90$+$n}{Other areas of general interest to physicists}
\Pacs{89}{70$+$c}{Information science}
\Pacs{05}{50$+$q}{Lattice theory and statistics; Ising problems}} 
\maketitle
\begin{abstract}
A variation of low density parity check (LDPC) error correcting codes
defined over Galois fields ($GF(q)$) is investigated using statistical
physics. A code of this type is characterised by a sparse random
parity check matrix composed of $C$ nonzero elements per column.  We
examine the dependence of the code performance on the value of $q$, for
finite and infinite $C$ values, both in terms of the thermodynamical
transition point and the practical decoding phase characterised by the
existence of a unique (ferromagnetic) solution. We find different
$q$-dependencies in the cases of $C=2$ and $C \ge 3$; the analytical
solutions are in agreement with simulation results, providing a
quantitative measure to the improvement in performance obtained using
non-binary alphabets.
\end{abstract}

Error correction mechanisms are essential for ensuring reliable data
transmission through noisy media.  They play an important role in a
wide range of applications from magnetic hard disks to deep space
exploration, and are expected to become even more important due to the
rapid development in mobile phones and satellite-based communication.

The error-correcting ability comes at the expense of information
redundancy.  Shannon showed in his seminal work~\cite{Shannon} that
error-free communication is theoretically possible if the code rate,
representing the fraction of informative bits in the transmitted
codeword, is below the channel capacity. In the case of unbiased
messages transmitted through a Binary Symmetric Channel (BSC), 
which we focus on here and which is characterized by a bit flip rate
$p$, the code rate $R=N/M$ which allows for an error-free transmission
satisfies
\begin{equation}
\label{eq:shannon_capacity}
R < 1-H_2(p), 
\end{equation}
where $H_2(p) \! = \! -p\log_2 p \! - \! (1-p)\log_2(1-p)$, when both
lengths of the original message $N$ and the codeword $M$ become
infinite.  The right hand side of (\ref{eq:shannon_capacity}) is often
termed {\em Shannon's limit}.

Unfortunately, Shannon's derivation is non-constructive and the quest
for practical codes which saturate this limit has been one of the
central topics in information theory ever since.  The current most
successful code in use is arguably the Turbo code~\cite{Turbo},
although the best performance to date, in terms of proximity to
Shannon's bound for a particular rate, has been achieved by variations
of the low-density parity check (LDPC) code, proposed by
Gallager~\cite{Gallager}.

A variation of Gallager's code has been recently discovered
independently by MacKay and Neal~\cite{MN}; an irregular construction
of this code, using a non-binary alphabet, provides the best error
correction performance to date~\cite{Davey}.  This discovery, based on
improving the code construction and the alphabet used by trial and
error, instigated the current work, aimed at clarifying the role
played by the alphabet used in obtaining this outstanding code
performance. To separate the effect of code irregularity from that of
the alphabet used we focus here on the dependence of {\em regular}
constructions on the chosen alphabet. To some extent this complements
our previous investigation on the impact of code irregularity on the
system's performance~\cite{us_JPA} in the case of binary alphabets.

Using a non-binary alphabet based on Galois fields $GF(q)$ is carried
out in the following manner: The sender first converts the Boolean
message vector $\mxi^B$ of dimensionality $N$ where $\xi_{i}^B\in
(0,1), \ \forall i$, to an $N/b$ dimensional vector of $GF(q=2^b)$
elements; where each segment of $b$ consecutive bits is mapped onto a
$GF(q)$ number\footnote{Binary vectors will be denoted by a
superscript B; other vectors are in the $GF(q)$ representation.}.  The
$GF(q)$ vector is then encoded to an $M/b$ dimensional $GF(q)$
codeword $\mz_0$, in the manner described below, which is then
reconverted to an $M$ dimensional Boolean codeword $\mz_0^B$,
transmitted via a noisy channel.  Corruption during transmission
can be modelled by the noise vector $\mzeta^B$, where corrupted bits
are marked by the value 1 and all other bits are zero, such that the
received corrupted codeword takes the form $\mz^B=\mz_0^B+\mzeta^B \
\mbox{(mod 2)}$.  The received corrupted Boolean message is then
converted back to a $GF(q)$ vector $\mz$, and decoded in the $GF(q)$
representation; finally the message estimate is interpreted as a
Boolean vector.

Firstly, we briefly explain the mapping of binary vectors onto the
Galois field $GF(q)$ elements.  
These represent a closed set of $q$ elements which can be added and 
multiplied utilizing an irreducible polynomial composed of 
Boolean coefficients. For instance, the irreducible polynomial for $GF(4)$ is 
$x^2+x+1$. Then, identifying the $b(=2)$ components of the binary vector with 
Boolean coefficients of a $b\!-\!1$ degree polynomial, 
$3 \oplus 1=(x+1) + 1 \ \mbox{(mod $2$)}
= x = 2 $ and 
$3 \otimes 2=(x+1) \times x \ \mbox{(mod $2$)}
= - 1 \ \mbox{(mod $2$)} = 1 $, setting $x^2+x+1=0 \ \mbox{(mod $2$)}$. 
Table I summarises the sum and product operations in Galois field $GF(4)$.
\begin{table}[t]
\begin{center}
\caption{Sum (left) and product (right) in $GF(4)$.}
\vspace*{1cm}
\begin{tabular}{c|c|c|c|c}
$\bigoplus$ & 0 & 1 & 2 & 3 \\
\hline
0 & 0 & 1 & 2 & 3 \\
\hline
1 & 1 & 0 & 3 & 2 \\
\hline
2 & 2 & 3 & 0 & 1 \\
\hline
3 & 3 & 2 & 1 & 0 \\
\hline
\end{tabular}
\hspace*{2cm}
\begin{tabular}{c|c|c|c|c}
$\bigotimes$ & 0 & 1 & 2 & 3 \\
\hline
0 & 0 & 0 & 0 & 0 \\
\hline
1 & 0 & 1 & 2 & 3 \\
\hline
2 & 0 & 2 & 3 & 1 \\
\hline
3 & 0 & 3 & 1 & 2 \\
\hline
\end{tabular}
\end{center}
\label{table:GF_q}
\end{table}
Secondly, we explain the encoding/decoding mechanism using regular
LDPC code in the $GF(q)$ representation. This is based on a randomly
constructed sparse parity check matrix $\A$ of dimensionality $(M-N)/b
\times M/b$.  This matrix is characterised by $C$ and $K$ nonzero
$GF(q)$ elements per column/row. The choice of $C$, $K$ is linked to
the code rate $R$, obeying the relation $C/K=1-R$.

Nonzero elements in each row are independently and randomly selected
from a specific distribution that maximises the entropy 
of the vector $\A\mzeta$ (all operations of vectors in the $GF(q)$
representation will be carried out as defined for this field; for
brevity we do not introduce different symbols to denote these
operations) when $\mzeta$ is the $GF(q)$ representation of the binary
random noise vector $\mzeta^B$. Then, one constructs an $M/b \times N/b$ 
generator matrix $\G^T$, which is typically dense, satisfying
$\A\G^T=0$~\cite{Davey}.

Using this matrix, encoding is carried out in the $GF(q)$
representation by taking the product $\mz_0=\G^T\mxi$; encoding is
performed by taking the product of the parity check matrix $\A$ and
the received corrupted message $\mz=\mz_0+ \mzeta$, which yields the
{\em syndrome} vector $\J = \A \mz = \A \mzeta$.  The most probable estimate of
the noise vector $\n$ is defined using the equation
\begin{equation}
\label{eq:decoding}
\A \n = \J \ , 
\end{equation}
via the iterative method of Belief Propagation (BP) ~\cite{MN}. This has 
been 
linked, in the case of Boolean codes, to the TAP (Thouless, Anderson,
Palmer) based solution of a similar physical system \cite{us_EPL}, a
relation which holds also in the case of $GF(q)$ codes as will be
shown elsewhere.

The noise vector estimate is then employed to remove the noise from
the received codeword and retrieve the original message $\mxi$ by
solving the equation $\G^T \mxi = \mz -\n $.

The similarity between error-correcting codes and physical systems was
first pointed out by Sourlas in his seminal
work~\cite{Sourlas_nature}, by considering a simple Boolean code, and
by mapping the code onto well studied Ising spin systems.  We recently
extended his work, which focused on extensively connected systems, to
the case of finite connectivity~\cite{us_EPL}.  Here, we generalise
these connections to spin systems in which the interaction is
determined using the $GF(q)$ algebra.

In order to facilitate the current investigation, we first map the
problem to that of a `$GF(q)$ spin system' of finite connectivity.
The syndrome vector $\J$ is generated by taking sums of the relevant
noise vector elements $J_\mu = A_{\mu i_1}\zeta_{i_1} + \ldots +
A_{\mu i_K}\zeta_{i_K} $, where $\mzeta=(\zeta_{i=1,\ldots,M/b})$
represents the true channel noise; the indices $i_1, \ldots, i_K$
correspond to the nonzero elements in $\mu$-th row of the parity check
matrix $\A=\left ( A_{\nu k} \right )$.  It should be noted that the
noise components $\zeta_i$ are derived from a certain distribution
$P_{pr}(\zeta_i)$, representing the nature of the communication
channel; this will serve as our prior belief to the nature of the
corruption process.  This implies that the most probable solution of
Eq.~(\ref{eq:decoding}) corresponds to the ground state of the
Hamiltonian
\begin{equation}
\label{eq:Hamiltonian}
{\cal H}(\n) \!=\! \sum_{\left \langle i_1, i_2, \ldots, i_K \right \rangle}
\! {\cal D}_{\left \langle i_1, i_2, \ldots, i_K \right \rangle} 
\left (1\!-\!\delta 
\left 
[J_{\left \langle i_1, i_2, \ldots, i_K \right \rangle};
A_{\mu i_1} n_{i_1}\! +\! \ldots \!+\! A_{\mu i_K} n_{i_K} \right] \right )
-\frac{1}{\beta}\sum_{i=1}^{M/b} \ln P_{pr}(n_i), 
\end{equation}
in the zero temperature limit $\beta = 1/T \to \infty$.  Elements of
the sparse tensor ${\cal D}_{\left \langle i_1, i_2, \ldots, i_K
\right \rangle}$ take the value $1$ if all the corresponding indices
of parity matrix $\A$ are nonzero in some row - $\mu$, and 0
otherwise.
The last expression on the right relates to the prior probability of
the noise vector elements. Note that operations between
vectors/elements in the $GF(q)$ representation (e.g., within the
$\delta$ function) are carried out as defined in this field.

The delta function provides $1$ if the contribution for the selected
site $A_{\mu i_1} n_{i_1} + \ldots + A_{\mu i_K} n_{i_K}$ is in
agreement with the corresponding syndrome value $J_{\left \langle
i_1, i_2, \ldots, i_K \right \rangle}$, recording an error, and $0$
otherwise.  Notice that this term is not frustrated as there are $M/b$
degrees of freedom while only $(M-N)/b$ constraints arise from
Eq.~(\ref{eq:decoding}), and its contribution can therefore vanish at
sufficiently low temperatures.  The choice of $\beta \to \infty$
imposes the restriction (\ref{eq:decoding}), limiting the solutions to
those for which the first term of (\ref{eq:Hamiltonian}) vanishes,
while the second term, representing the prior information about the
noise, survives.

The optimal estimator, minimising the expectation of discrepancy per
noise bit, is of the form $\widehat{n}_i= \mbox{argmax}_{a \in {GF}(q)}
\left \langle \delta(n_i,a) \right \rangle_{\beta \to \infty}$.  This
is known as the {\em marginal posterior maximiser} (MPM)~\cite{Iba}
and corresponds to the finite temperature decoding at Nishimori's
temperature studied in other
codes~\cite{Sourlas_EPL,Rujan,Nishimori_code}.  Notice that here, due
to the hard constraints imposed on the dynamical variables, decoding
at zero temperature is optimal, as the true posterior distribution
(given $\J$) relates to the ground state of Hamiltonian
(\ref{eq:Hamiltonian}), similar to other LDPC codes~\cite{us_PRL}.  The
macroscopic quantity $m=(b/M)\left \langle \sum_{i=1}^{M/b}
\delta(\widehat{n}_i,\zeta_i) \right \rangle_{\{{\cal D},A,\mzeta\}}$
serves as the performance measure.

To eliminate the dependency of the syndrome $J_{\left \langle i_1,
\ldots, i_K \right \rangle }$ on the noise vector $\mzeta$ we employ
the gauge transformation $n_i \to n_i + \zeta_i$, $J_{\left \langle
i_1, \ldots, i_K \right \rangle } \to 0$.  Rewriting
Eq.~(\ref{eq:Hamiltonian}) in this gauge moves the dependency on
$\mzeta$ to the second term where it appears in a decoupled form
$(1/\beta)\ln P_{pr}(n_i + \zeta_i)$.  The remaining difficulty comes
from the complicated site dependency caused by nontrivial $GF(q)$
algebra in the first term.  However, one can rewrite this dependency
in a simpler form 
\begin{eqnarray}
\label{eq:products}
\delta
\left [0;A_{\mu i_1} n_{i_1} + \ldots + A_{\mu i_K} n_{i_K} \right]
&=& \!\!\!\!\! \!\!\!\!\! \!\!\!\!\! \!\!\!\!\!
\sum_{A_1,\ldots,A_K,a_1, \ldots,a_K =0} ^{q-1} \!\!\!\!\! \!\!\!\!\! 
\delta \left [0;A_{1} a_{1} + \ldots + A_{K} a_{K} \right] \\
&\times& \!\!\!\!\! 
\delta(A_1,A_{\mu i_1} ) \ldots \delta(A_K,A_{\mu i_K}) 
\times 
\delta(a_1,n_{i_1}) \ldots  \delta(a_K,n_{i_K}) \ ,  \nonumber
\end{eqnarray}
by introducing Kroncker's $\delta$ and the
dummy variables $A_1,\ldots,A_K$ and $a_1, \ldots, a_K$. 

Since codes of this type are usually used for long messages with
$N=10^3-10^5$, it is natural to resort to the methods of statistical
mechanics for analysing their properties.  The random selection of
sparse tensor ${\cal D}$, identifying the nonzero elements of $\A$,
and the noise vector $\mzeta$ introduces quenched disorder to the
system. More specifically, we calculate the partition function ${\cal
Z} \left ( {\cal D}, A, \mzeta \right )= \mbox{Tr}_{\n} \exp \left [
-\beta {\cal H} \right ]$ averaged over the disorder and the
statistical properties of the noise estimation, using the replica
method~\cite{us_EPL,us_PRL}.  Taking $\beta \to \infty$ gives rise to
a set of order parameters
\begin{eqnarray}
\label{eq:orderparameters}
{\cal Q}_{a_1,a_2,\ldots,a_n}=\frac{b}{M}
\sum_{i=1}^{M/b}  \left \langle Z_i \prod_{\alpha=1}^n
\left \langle \delta \left (a_\alpha,n_{i \alpha} \right ) 
\right \rangle_{\beta \to \infty}
\right \rangle_{{\cal D}, A, \mzeta}, 
\end{eqnarray}
where $\alpha=1,\ldots,n$ represents the replica index and $a_\alpha$
runs from $0$ to $q-1$, and the variables $Z_i$ come from enforcing
the restriction of $C$ connections per index $i$
\begin{equation}
\delta \left ( 
\sum_{
\left \langle 
i_2,\ldots,i_K 
\right \rangle}
{\cal D}_{
\left \langle i,i_2,\ldots,i_K \right \rangle}
-C \right )
= \oint \frac{dZ}{2 \pi}
Z^{
\sum_{
\left \langle 
i_2,\ldots,i_K 
\right \rangle}
{\cal D}_{
\left \langle i,i_2,\ldots,i_K \right \rangle}
-(C+1)}. 
\end{equation}

To proceed further, one has to make an assumption about the symmetry of 
order parameters. The assumption made here is that of replica symmetry 
reflected in the representation of the order parameters and of the 
related conjugate variables:
\begin{eqnarray}
\label{eq:rs_ansatz}
{\cal Q}_{a_1,a_2,\ldots,a_n}&=&a_{\cal Q} \int d \PP \ 
\pi(P_0,\ldots,P_{q-1})
\prod_{\alpha=1}^n P_{a_\alpha}, \cr
\widehat{\cal Q}_{a_1,a_2,\ldots,a_n}&=&a_{\widehat{\cal Q}} \int d 
\widehat{\PP } \ 
\widehat{\pi}(\widehat{P}_0,\ldots,\widehat{P}_{q-1})
\prod_{\alpha=1}^n \widehat{P}_{a_\alpha}, 
\end{eqnarray}
where $a_{\cal Q}$ and $a_{\widehat{\cal Q}}$ are normalisation
coefficients; $\pi(\PP)$ and $\widehat{\pi}(\widehat{\PP})$ represent
probability distributions for $q$ dimensional vectors
$\PP=(P_0,\ldots,P_{q-1})$ and
$\widehat{\PP}=(\widehat{P}_0,\ldots,\widehat{P}_{q-1})$,
respectively.  Unspecified integrals are performed over the region
$P_0+\ldots+P_{q-1}=1$, $P_{a=0,\ldots,q-1} \ge 0 $ or
$\widehat{P}_0+\ldots+\widehat{P}_{q-1}=1$,
$\widehat{P}_{a=0,\ldots,q-1} \ge 0 $.  Extremising the averaged
expression with respect to the probability distributions, one obtains
the following free energy per spin
\begin{eqnarray}
\label{eq:free_energy}
- \frac{b}{M} &\phantom{=}& \hspace*{-2em}\left \langle 
\ln {\cal Z} \right \rangle_{{\cal D},A,\mzeta}
  = -
\mathop{\rm Ext}_{\{\pi,\widehat{\pi}\}} 
\left \{ \int \prod_{l=1}^C d \widehat{\PP}^l \ \widehat{\pi} \left (\widehat{\PP}^l 
\right ) 
\left \langle 
\ln \left (\sum_{a=0}^{q-1} \prod_{l=1}^C 
\widehat{P}_a^l \ P_{pr} \left (a + \zeta \right ) \right )
\right \rangle_\zeta \right . \cr
 &+&
\frac{C}{K}\int \prod_{l=1}^K d \PP^l \pi \left (\PP^l \right ) 
\left \langle 
\ln \left ( 
\sum_{a_1,\ldots, a_K=0}^{q-1}
\delta \left [0;A_{1} a_{1} + \ldots + A_{K} a_{K} \right]
\prod_{l=1}^K P_{a_l}^l
\right )
\right \rangle_A  \cr
&-& \left . C \int d \PP  \ d \widehat{\PP} \
\pi\left (\PP \right ) \
\widehat{\pi}  (\widehat{\PP}  ) 
\ln \left ( \sum_{a =0}^{q-1} P_a \widehat{P}_a \right ) 
 \right \} \ , 
\end{eqnarray}
where $\left \langle \cdot \right \rangle_A$ and $\left \langle \cdot
\right \rangle_\zeta$ denote averages over the distribution of nonzero
units per row in constructing the matrix $\A$ and over $P_{pr}(\zeta)$,
respectively.

One can calculate the free energy via the saddle point method. 
Solving the equations obtained by varying 
Eq.(\ref{eq:free_energy}) is generally difficult. 
However, it can be shown analytically that 
a successful solution 
\begin{eqnarray}
\label{eq:success}
\pi (\PP)= \delta(P_0 -1) \prod_{a =1}^{q-1}
\delta(P_a), \quad 
\widehat{\pi} (\widehat{\PP})= \delta(\widehat{P}_0 -1) 
\prod_{a =1}^{q-1}
\delta(\widehat{P}_a), 
\end{eqnarray}
which implies perfect decoding $m=1$, 
extremises the free energy for $C \ge 2$. 
For $C \to \infty$, an unsuccessful solution, which 
provides $m < 1$, is also obtained analytically 
\begin{eqnarray}
\label{eq:unsuccess}
\pi (\PP)= \left \langle \prod_{a=0}^{q-1}
\delta 
\left ( P_a -P_{pr}(a + \zeta ) \right )
\right \rangle_{\zeta}, 
\quad 
\widehat{\pi} (\widehat{\PP})= \prod_{a=0}^{q-1}
\delta 
\left ( \widehat{P}_a - \frac{1}{q} \right ). 
\end{eqnarray}
Inserting these solutions into (\ref{eq:free_energy}) it is found that
the solution (\ref{eq:success}) becomes thermodynamically dominant
with respect to (\ref{eq:unsuccess}) for $R < 1-H_2(p)$ independently
of $q$; which implies that the code saturates Shannon's limit for $C
\to \infty$ as is reported in the information theory
literature~\cite{Davey}.

Finding additional solutions analytically is difficult, we therefore
resorted to numerical methods.  Approximating the distributions
$\pi(\PP)$ and $\widehat{\pi}(\widehat{\PP})$ by $5 \times 10^3 - 3
\times 10^4$ sample vectors of $\PP$ and $\widehat{\PP}$ we obtained
solutions by updating the saddle point equations ($100 - 500$
iterations) for codes of connectivity $C=2, \ldots, 6$ and $GF(q)$
representation $q=2,4,8$ and for both BSC and Gaussian channels. Less
then $50$ iteration were typically sufficient for the solutions to
converge. Due to lack of space we present here results only for the
case of the BSC; results for the case of Gaussian channels are
qualitatively similar and will be presented elsewhere.

Since the suggested properties are different for $C \ge 3$ and $C =
2$, we describe the results separately for the two cases.  For $C \ge
3$, it turns out that Eq.(\ref{eq:success}) is always locally
stable. However, an unsuccessful solution, approaching 
(\ref{eq:unsuccess}) as $C \to \infty$, becomes thermodynamically
dominant for sufficiently large flip rate $p$.  As the noise level is
reduced, the solution (\ref{eq:success}) becomes thermodynamically
dominant at a certain flip rate $p=p_t$, and remains dominant until $p
\to 0$. This implies that perfect decoding $m=1$ is feasible for
$p<p_t$.  However, the locally stable 
unsuccessful solution remains as well above a 
certain noise level $ p_s (\le p_t)$.

As $C \to \infty$, the transition point $p_t$ converges from below
to Shannon's limit $p_c=H_2^{-1}(1-R)$ irrespective of the value of
$q$.  For finite $C$, $p_t$ monotonically increases with $q$ but does
not saturate $p_c$. This implies that error correcting ability of the
codes when optimally decoded is monotonically improved as $q$
increases.

The behaviour of the spinodal point $p_s$ is quite different, as shown
in Fig.1a, presenting the dependence of $p_t$ and $p_s$ on $q$ for
connectivity $C=4 \ (\ge 3)$. It appears that $p_s$ is generally
decreasing with respect to $q$ (except for unique pathological cases),
indicating a lower practical corruption limit for which BP/TAP
decoding will still be effective. Above this limit BP/TAP dynamics is
likely to converge to the unsuccessful solution due to its dominant
basin of attraction~\cite{us_PRL}.
\begin{figure}
\label{fig:fig1}
\begin{center} 
\leavevmode  
\epsfysize =5.5cm 
\epsfbox{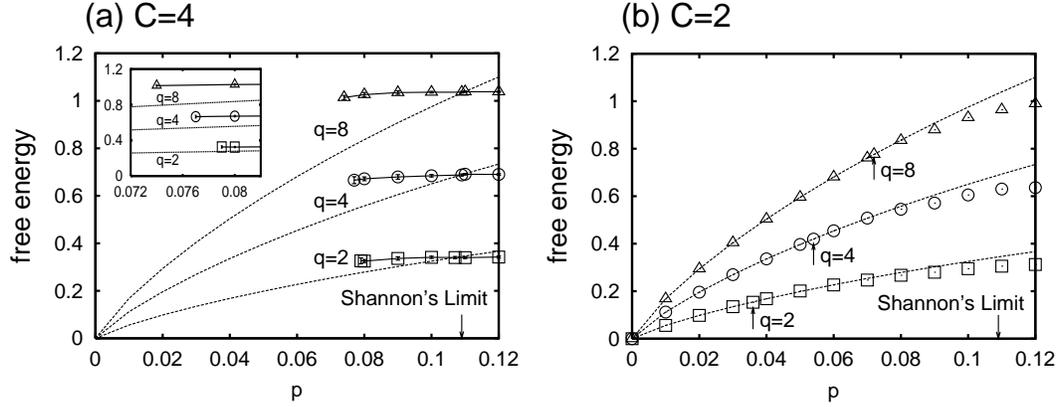}
\end{center} 
\caption{Extremised free energies (\ref{eq:free_energy}) obtained for
$q=2,4,8$ as functions of the flip rate $p$ (a) for connectivity $C=4 \
(\ge 3)$ and (b) for $C=2$ codes with the same code rate
$R=1-C/K=1/2$.  In both codes, broken lines represent free energies of
successful solution (\ref{eq:success}), while markers stand for
unsuccessful solutions ($m < 1$).  Monte Carlo methods with $5 \times
10^3 - 3 \times 10^4$ samplings at each step are employed for
obtaining the latter with statistical fluctuations smaller than the
symbol size.  For each $q$ value, the solution having the lower free
energy becomes thermodynamically dominant.  In (a), crossing points
provide the critical flip rate for $p_t$ being 
$0.106, 0.108$ and $0.109$ (within the numerical precision) 
for $q=2,4$ and $8$, respectively, monotonically approaching
Shannon's limit $p_c=H^{-1}(1/2)=0.109$.  The inner figure focuses on
the vicinity of the spinodal points $p_s$, determining the limit of
successful practical decoding.  This shows $p_s$ to decrease with
increasing $q$.  (b) shows that $C=2$ codes exhibit continuous
transitions between the successful and unsuccessful solutions.  The
critical flip rate $p_{b}$, pointed by arrows, is increases with $q$,
while it is still far from Shannon's limit. }
\end{figure}
In contrast, $C=2$ codes exhibit a different behaviour; the solution
(\ref{eq:success}) becomes the unique minimum of free energy
(\ref{eq:free_energy}) for sufficiently small noise levels, which
implies that practical decoding dynamics always converges to the
perfect solution. However, as the noise level increases, the solution
loses its stability and continuously bifurcates to a stable suboptimal
solution.  Unlike the case of $C \ge 3$, this bifurcation point $p_b$,
which monotonically increases with $q$, determines the limit of
practical BP/TAP decoding.  The practical limit obtained is
considerably lower than both Shannon's limit and the thermodynamic
transition point $p_t$ for other $C \ge 3$ codes with the same $q$
value (Fig. 1b).  Therefore, the optimal decoding performance of $C=2$
codes 
is the worst within this family of codes.

However, $p_b$ can become closer to, and even higher than, the spinodal
point $p_s$ of other $C \ge 3$ codes for large $q$ values, (Table II)
implying that the practical decoding performance of $C=2$ codes is not
necessarily inferior to that of $C \ge 3$ codes. 
This is presumably due to the decreasing solution numbers 
to Eq.(\ref{eq:decoding}) for $C=2$ as $q$ increase, compared to 
the moderate logarithmic increase in the information content, 
tipping the balance in favour of the perfect solution.
This may shed light on the role played by $C=2$ elements 
in irregular constructions.
\begin{table}[ht]
\begin{center}
\caption{The critical noise level, below which BP/TAP-based decoding
works successfully, for different connectivity values $C$ in the case
of $q=8$ and $R=1-C/K=0.5$.  This is determined as the spinodal point
$p_s$ and the bifurcation point $p_b$ for $C \ge 3$ and $C=2$,
respectively.  The critical noise for $C=2$ becomes higher than that
of $C \ge 5$. }
\vspace*{0.2cm}
\begin{tabular}{c|ccccc}
$C$ & $2$ & $3$ & $4$ & $5$ & $6$ \\
\hline
Critical noise & 0.072 & 0.088 & 0.073 & 0.062 & 0.050 \\
\hline
\end{tabular}
\end{center}
\end{table}

In summary, we have investigated the properties of LDPC codes defined
over $GF(q)$ within the framework of statistical mechanics.  Employing
the replica method, one can evaluate the typical performance of codes
in the limit of infinite message length.  It has been shown
analytically that codes of this type saturate Shannon's limit as $C
\to \infty$ irrespective of the value of $q$, in agreement with
results reported in the information theory literature~\cite{Davey}.
For finite $C$, numerical calculations suggest that these codes
exhibits two different behaviours for $C \ge 3$ and $C=2$.  For $C \ge
3$, we show that the error correcting ability of these codes, when
optimally decoded, is monotonically improving as $q$ increases;
while the practical decoding limit, determined by the emergence of a
suboptimal solution, deteriorates.  On the other hand, $C=2$ codes
exhibit a continuous transition from optimal to sub-optimal solutions
at a certain noise level, below which practical decoding dynamics
based on BP/TAP methods converges to the (unique) optimal
solution. This critical noise level monotonically increases with $q$
and becomes even higher than that of some codes of connectivity $C
\ge 3$, while the optimal decoding performance is inferior to that of
$C \ge 3$ codes with the same $q$ value. This may elucidate the role
played by $C=2$ components in irregular constructions.

Future directions include extending the analysis to irregular Gallager
codes as well as to regular and irregular MN code~\cite{MN,us_PRL} in
the Galois representation.

\stars

We acknowledge support from the RFTF program of the JSPS (YK), EPSRC
(GR/N00562) and The Royal Society (DS).

\end{document}